\documentclass[aps,pre,reprint,superscriptaddress]{revtex4-1}

\usepackage{amsmath}
\usepackage{hyperref}
\usepackage{graphicx}

\graphicspath{{data/}{data/equilibration/}{data/corr_complete/}{data/corr_ER/}}

\begin{document}

\title{Large Deviation Properties of Minimum Spanning Trees for Random Graphs}

\author{Mahdi Sarikhani}
\affiliation{Department of Physics, College of Science, Shiraz University, Shiraz, Iran}

\author{Alexander K. Hartmann}
\affiliation{Institute of Physics, University of Oldenburg, Oldenburg, Germany}

\date{\today}

\begin{abstract}
    We study the large-deviation properties of minimum spanning trees
    for two ensembles of random graphs with $N$ nodes.
    First, we consider complete graphs. Second, we study
    Erd\H{o}s-R\'{e}nyi (ER) random graphs with
    edge probability $p=c/N$ conditioned to be
    connected. By using large-deviation Markov
    chain sampling, we are able to obtain the distribution $P(W)$ of
    the spanning-tree weight $W$ down to probability densities as
    small as $10^{-300}$.
    For the complete graph, we confirm analytical predictions
    with respect to the expectation value. For both ensembles, the
    large deviation principle is fulfilled. For the connected ER
    graphs, we observe a remarkable change of the distributions at the
    value of $c=1$, which is the percolation threshold for the original
    ER ensemble.
\end{abstract}

\maketitle

\section{Introduction}

Graphs and networks
\cite{albert2002review,newman2003review,boccaletti2006,dorogovtsev_book2006,newman_book2010,barrat_book2012},
are ubiquitous in modeling systems like chemical molecules,
disordered materials, energy grids, spread of diseases, or social networks.
Graphs can be analyzed in many ways, like the calculation of the
connected components, graph diameter, resilience to node or edge
removal, communities,
minimum matchings, spectral properties, and many more; all these
have found applications in physical systems.

Here we consider
\emph{spanning trees}, which are subsets of edges of a graph
that connect all vertices.
Being a tree, for $N$ nodes the spanning tree has $N-1$ edges, and
from each vertex to any other one, there exists a unique path. If the
edges carry weights, like distances or costs, the
\emph{minimum spanning tree}
is among all spanning trees the one where the sum $W$ of
the weights of the tree edges is a minimum.

The study of optimum trees dates back \cite{graham1985}
to the works of Fermat and Torricelli in the 17th century.
In practical applications and research, spanning trees have found many
applications, e.g., to
set up transportation networks \cite{prim1957},
reconstruct evolutionary trees \cite{cavalli-sforza1967},
cluster data \cite{gower1969}, detect communities \cite{wu2013},
determine patterns in images \cite{osteen1974}, and many more.
In statistical physics, among several other applications,
spanning trees have been analyzed by counting \cite{zhang2012}
or enumerating \cite{li2019} them.
Also, spanning trees have been
used to analyse atomic deposition structures on surfaces \cite{dussert1986},
avalanches in sandpile models \cite{ktitarev2000},
transport \cite{wu2006} and  other technical  networks \cite{kim2004},
stock markets \cite{bonanno2003,onnela2003}, and fractal properties of
percolation clusters \cite{jackson2010a,jackson2010b,sweeny2013}.
Finally, spanning trees have also attracted the attention of mathematicians,
where we use  some of the work
\cite{fenner1982,frieze1985,mcDiarmid1989,janson1995,flaxman2007,cooper2016}
for comparison with our results below.

In this work, we study numerically the distribution $P(W)$ of the weights of
minimum spanning trees for two random ensembles. The first one consists
of complete graphs, where the edge weights are randomly,
independently and identically
distributed (iid) uniformly in the interval $[0,1]$. For this case,
the expectation value of $W$ in the limit of a large number $N$ of nodes is known
\cite{frieze1985}, also that the probability to deviate
from this value decreases exponentially with $N$. But details
about the distribution are not known. The second ensemble we consider
consists of Erd\H{o}s-R\'{e}nyi (ER) random graphs conditioned to be
connected, again the weights are iid uniform in $[0,1]$.
Here, to the best of our knowledge, nothing is known about the
distribution $P(W)$. The approach we use utilizes a Markov-chain
Monte Carlo large-deviation sampling \cite{bucklew2004, largest-2011}
of random graphs, which allows
us to access the tails of the distributions down to exponentially small
probabilities such as $10^{-300}$. This means we can obtain
the distribution over a very large range of support. This ability is useful
to verify the so-called large-deviation principle
\cite{denHollander2000,touchette2009}.
Also, we hope that our study motivates more analytical work,
possibly involving approximation techniques, which can be compared
to our results.

The paper is organized as follows. In Sec.~\ref{sec:model} we introduce
minimum spanning trees, the graph ensembles we use, the
quantities we measure, and mention previous
results, which are relevant for our work.
Next, we explain the algorithms we have used to obtain the
distribution of spanning-tree weights over hundreds of decades in
probability density.
In Sec.~\ref{sec:results} we present our results before we finish with
a summary and outlook.

\section{Models\label{sec:model}}

A \emph{graph} $ G = (V, E) $ consists of $ N $ nodes $ i \in V $
and $M$ undirected edges $ \{ i, j \} \in E \subset V^{(2)} $.
For each edge $ \{ i, j \} \in E $, the nodes $ i $ and $ j $
are called \emph{adjacent}.
The edge is said to be \emph{incident} to these two adjacent nodes.
The \emph{degree} of a node $ i \in V $ is the number of nodes
adjacent to $i$.
A sequence $P=(i_1,\ldots, i_l)$ of nodes, where all pairs $i_{k},i_{k+1}$ of
consecutive nodes are adjacent, is called a \emph{path} of length $l-1$;
the two nodes $i_1$ and $i_l$
are called \emph{connected} by the path $P$.
If a path connects a node with itself,
i.e., $i_1=i_{l}$, the path is called a \emph{cycle}. A graph that does
not contain a cycle is called \emph{acyclic}.
If in a graph all nodes are connected, the
graph is also called connected.
Here we consider \emph{weighted} graphs, i.e.,
each edge $ \{ i, j \} \in E $ is assigned a real-valued
weight $ w_{i,j} $.

A \emph{spanning tree} $ T=(V,E_T) $ for a given graph $G=(V,E)$
consists of the set $V$ of nodes
and an acyclic subset $E_T$ of $ E $ that
connects all nodes, i.e., a tree that spans the graph $ G $.
The total edge weight of a spanning tree is the sum
\begin{equation}
    \tilde W\equiv \tilde W(T) = \sum_{\{ i, j \} \in E_T} w_{i,j}
\end{equation}
of the weights of the edges of the tree.
The \emph{minimum spanning tree} (MST) of graph $ G $
is a spanning tree where the weight $\tilde W$ is minimal;
we denote the minimum weight as $W$.
Note that in
general the minimum spanning tree may not be unique, but for the
ensembles we consider here this is the case, so we can speak of
\emph{the} MST, and we also write $W(G)$ for $W(T(G))$.

Here we do not consider just single specific given graphs,
but ensembles of a fixed or random
connected graphs with, in any case, randomly chosen edge weights,
which means that the minimum spanning-tree weight $W$
is a random variable and will be described by a probability distribution
with density $P(W)$.
If we denote the probability density to obtain a certain graph $G$,
which is meant to include the edge weights, by $Q(G)$, one has
\begin{equation}
    P(W) = \sum_{G} \delta_{W(G), W} Q(G)\,.
\end{equation}
Note that since the weights are continuous variables, the $\sum_G$
actually involves also an integral, but we keep the summation
for simplicity.
Obtaining, i.e., numerically estimating $P(W)$, including the low-probability
tails, is the purpose of this work.

The first ensemble we consider consists of weighted complete graphs
$G=(V, V\times V)$, where only the weights $w_{i,j}$
are uniform iid random variables
in $[0,1]$, i.e.,
$$
    w_{i,j} \sim U(0,1).
$$

For this ensemble, the expectation value $\langle W \rangle_N$ was obtained
analytically \cite{frieze1985}
in the limit of infinite graph size $N\to \infty$
as
\begin{equation}
    \langle W \rangle_N \to \zeta(3)/D\,\, \text{where} \;
    \zeta(3)=\sum_{k=1}^\infty 1/k^3 = 1.202\ldots\,.
    \label{eq:expectation}
\end{equation}
This holds actually for
arbitrary distribution functions $F(w_{ij})$ of
\emph{non-negative} edge weights,
and $D=F'(0)>0$ is the probability density at 0. For the present case
with uniform distribution in $[0,1]$, we have $D=1$. This is compatible
with a previous result \cite{fenner1982} that
$\langle W \rangle_N \le 2(1+\log n /n)$.
Recently, also the finite-size corrections of $\langle W \rangle_N$
have been calculated \cite{cooper2016}.
With respect to the distribution,
the typical part becomes Gaussian in the limit of large $N$ with variance
about 1.6857 \cite{janson1995}.
Concerning the tails, which is our main concern here,
it has been shown \cite{mcDiarmid1989,flaxman2007}
that the deviations from the typical
value decrease exponentially in the graph size $N$, i.e., fulfill
the large-deviation principle; see the end of this section.

Second, we study
a special ER \cite{erdoes1960} random graph ensemble.
Standard ER random graphs with parameter $p\in[0,1]$
are constructed as follows. One
starts with a set of $N$ vertices. Then one
iterates over the $ N (N - 1) / 2 $ possible pairs of nodes and
adds each edge $ \{ i, j \} \in V^{(2)} $ with probability $p$.
Of particular interest are sparse ER graphs where $p=c/N$ and
$ c $ denotes the average degree of a node in the graph.

To study MSTs, we require the graphs to be connected.
Thus, the second ensemble consists of ER random graphs conditioned
to be connected. Technically, depth-first search \cite{cormen2001}
is used to verify connectiveness. If a graph is not connected,
it is discarded in the analysis. Note that for small values of $c$,
typical ER graphs will not be connected. Here the Markov chain Monte
Carlo (MCMC) approach presented below is more efficient to
generate connected ER random graphs.

In large-deviation
theory \cite{denHollander2000,touchette2009}, one considers
the probabilities $ P(X) $ for
random quantities $X$. The tails of $P(X)$ describe
the large deviations from the typical values of $X$. Here we
consider $X=W$, and the corresponding intensive quantity $w=W/(N-1)$.
Often these deviations are for large values of $N$
exponentially small in $ N $ as
\begin{equation}
    P(W) \simeq e^{-N \Phi(w)}\,,
    \label{eq:rate-function}
\end{equation}
with the \emph{rate function} $\Phi(w)$ which often depends
on the intensive quantity. This form of the distributions
has a very specific dependence on the system size $N$, separated
from the intense quantity $w$.
Using the small-$ o $ notation, this means that
\begin{equation}
    P(W) = e^{-N \Phi(w) + o(N)}, \quad (N \to \infty).
\end{equation}
The so-called
\emph{large deviation principle} holds if, loosely speaking, the
distribution has shape Eq.~(\ref{eq:rate-function})
and the empirical
rate function
\begin{equation}
    \label{eq:rate_function}
    \Phi_{N}(w) \equiv - \frac{1}{N} \ln P(wN),
\end{equation}
converges to $ \Phi(w) $ as $ N \to \infty $. Due to the logarithm,
both the normalization and the subleading term of $ P(W) $
become additive contributions to $ \Phi $, which go to zero
as $ N \to \infty $.

\section{Algorithms}

An MST can be conveniently calculated by using
Prim's algorithm \cite{prim1957,cormen2001}.
It begins with a randomly selected node
$i$ which is the first node of the tree. Then the algorithm
treats $i$ and all subsequent
nodes added to the tree in the same way: all edges adjacent
to it are added to a priority queue.
Iteratively the current minimum-weight edge $\{k,l\}$
is pulled out of the queue. At least one node of the edge
will already belong to the tree.
It is checked whether even both nodes $k$ and $l$ belong to the tree.
If yes, the edge is disregarded.
If not, say $k$ is in the tree but not $l$; $l$ is added
to the set of tree nodes.
Also the edge $\{k,l\}$ is added to the tree edges and the other
edges incident to $l$, i.e., except $\{k,l\}$,
 are inserted into the priority queue. When, during the selection
from the priority queue, $ m $ edges with the equal
(smallest) weight appear, the degeneracy is broken by randomly
selecting one with probability $ 1 / m $. As mentioned above, due
to the real-valued edge weights, there will be no degeneracy here.
The algorithm stops when all nodes belong to the tree.
The running time of Prim's algorithm as described above is of the order
of $O(M \log N)$.

To estimate the distribution $ P(W) $ in the high probability region,
direct sampling is straightforward:
one generates $ K $ graph samples and determines the weight of the
minimum spanning tree $ W(G) $ for each sample $ G $. Each graph
$ G $ occurs with its natural ensemble probability density $ Q(G) $.
Therefore, by calculating a histogram of the values for $W$,
an estimation for $ P(W) $ can be obtained.
Also, the mean or the variance can be estimated rather accurately.
Nevertheless, with this
direct sampling, $ P(W) $ can only be measured in a regime where $ P(W) $
is relatively large, about $ P(W) > 1 / K $. Unfortunately,
the distribution usually decays exponentially in the system size $ N $
when moving away from its typical (peak) value. Thus, even for
moderate system sizes, the distribution remains unknown over most of
its support.

To estimate $ P(W) $ for a wider range of weights of the minimum
spanning trees, including very small probability densities
of the order of $ 10^{-100} $, we use
a different approach \cite{align2002,bucklew2004,largest-2011}.
Here we display only the main ingreditents of the algorithm,
for a pedagogical introduction, see Ref.~\cite{les_houches2024}.
The basic idea is to generate graphs with a modified density
that includes an additional exponential (Boltzmann)
factor $ \exp(-W(G) / \theta ) $,
where $ \theta $ is a temperature-like parameter,
which controls the weight.

To sample according to the original weights modified by the additional
Boltzmann factor, we
perform standard MCMC simulations,
with the current state at step $ t $ is given by an instance $ G(t)$ of a
graph. The Metropolis-Hasting algorithm
\cite{metropolis1953,hastings1970,newman1999,landau2000}
is applied as follows.
At each step $ t $, a candidate graph $ G^{*} $ is created from
the current graph $ G(t) $. Different techniques to create candidate
graphs are used for different graph ensembles:
\begin{itemize}
    \item
          In the case of complete graphs, an edge $ \{ i, j \} $ is selected
          randomly,
          with uniform probability $ 1 / M $, where $ M = N (N - 1) / 2 $.
          The weight of the selected edge is changed to a new random value with uniform
          probability in the range $ [0,1] $.
    \item
          In the case of ER random graphs, a node $ i $ is selected randomly
          with uniform probability $ 1 / N $. All edges adjacent
          to $ i $ are removed. Then, for all pairs $ i, j $,
          new edges $\{i,j\}$
          are added with probability $ p = c / N $,
          and their weights are drawn uniformly from the range $ [0,1] $,
          respectively.
\end{itemize}

If the candidate graph $ G^{*} $ is not connected, it will be immediately
rejected.
Note that the initial graphs also need to be connected.
Therefore, most of our simulation begins with a complete graph, and
the described procedure runs until the equilibration is reached; see below.

If a connected candidate graph $ G^{*} $ is created, we calculate the
weight of its minimum spanning tree $ W(G^{*}) $.
The candidate graph is accepted ($ G(t + 1) = G^{*} $) with the
Metropolis probability
\begin{equation}
    p = \min \left\{ 1, e^{-[W(G^{*}) - W(G(t))] / \theta} \right\}.
\end{equation}
In case the candidate graph is rejected, the
current graph is kept, i.e., $ G(t + 1) = G(t) $.

The temperature parameter $ \theta $ allows us to bias the
sampling toward graphs with larger or smaller MST weights.
An infinite temperature leads to acceptance of all configurations,
which gives access to the typical values. Positive (negative)
$ \theta $ allows us to sample graphs with smaller (larger) weights
of the MST than the typical one.

\begin{figure}
    \includegraphics[width=0.99\columnwidth]{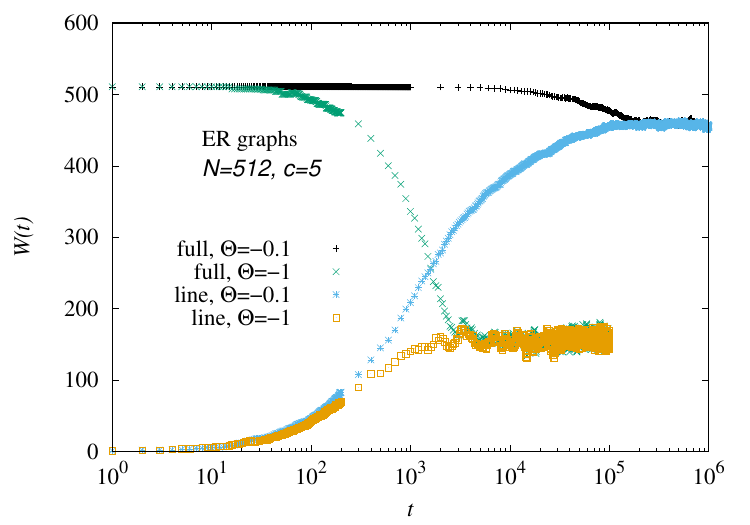}
    \caption{\label{fig:equilibration}
        Equilibration of Markov chain: time series $W(t)$ for ER graphs
        with $N=512$ and $c=5$ for two different
        initial configurations with full graphs (all $w_{ij}=1$) and a line graph
        (all $w_{ij}$ near 0.)}
\end{figure}

By construction, the algorithm fulfills detailed balance. It is also ergodic,
since within $M$ or $N$ steps, respectively,
each possible graph may be constructed.
To check equilibration, we monitor the time series $W(t)$. The most simple
idea is to consider the time series equilibrated if no systematic trend
is visible anymore, only fluctuations. This is in particular well visible
if one starts two independent runs
with initial configurations that lead to very different
initial weights $W(0)$, respectively, such that the time series will converge
to the equilibrium range from different sides. As an example, we
consider the ER ensemble with $N=512$ and $c=5$. For one initial graph,
we take the complete graph with all weights set to $w_{ij}=1$. Thus $W(0)=511$.
For the other initial graph, we start with a line graph of 511 edges where all
weights are randomly close to zero. Here we have $W(0)$ about 50. The
resulting time series $W(t)$ are shown \footnote{The data is publicly
available in the repository DARE of the University of Oldenburg
\cite{data_spanning_tree}.}
in Fig.~\ref{fig:equilibration}
for two different temperatures, $\theta=-0.1$ and $\theta=-1$. For the
first case, equilibration takes some time, more than $10^5$ steps,
while for $\theta=-1$, equilibration is reached faster.
Note that for $\theta=-0.1$,
the observed weights are close to the maximum possible value $W=512$, and
with small fluctuations. So for the simulations to obtain the
distributions, we did attempt to reach
such extreme values of the weight, i.e., the far tails, by suitably chosen
temperature values.

Thus, in the limit of infinitely long Markov chains, the distribution of graphs
will follow the probability
\begin{equation}
    Q_{\theta}(G) = \frac{1}{Z(\theta)} e^{-W(G) / \theta} Q(G),
\end{equation}
where $Q(G)$ is the above introduced original probability of the
graph within its ensemble and
$ Z(\theta) $ is an a priori unknown normalisation constant.

The probability distribution for $ W $ at temperature $ \theta $ is given by,
where the $\sum_G$ is meant to involve
again an integral over the possible weight
values for all edges:
\begin{align}
    P_{\theta}(W) & = \sum_{G} \delta_{W(G), W} Q_{\theta}(G) \nonumber                          \\
                  & = \frac{1}{Z(\theta)} \sum_{G} \delta_{W(G), W} e^{-W(G) / \theta} Q(G)
    \nonumber                                                                                    \\
                  & = \frac{e^{-W / \theta}}{Z(\theta)} \sum_{G} \delta_{W(G), W} Q(G) \nonumber
    \\
                  & = \frac{e^{-W / \theta}}{Z(\theta)} P(W) \nonumber                           \\
    \implies P(W) & = e^{W / \theta} Z(\theta) P_{\theta}(W).
    \label{eq:probability_distribution}
\end{align}

Thus, the target distribution $ P(W) $ can be estimated from
$ P_{\theta}(W) $ at finite temperatures, up to a normalization constant
$ Z(\theta) $. Note that for a finite sample of graphs, each selected
temperature $\theta$ will lead to an observation of
values of $W$ within a specific range
and will therefore allow us to estimate $P(W)$ there.
A bias toward smaller (larger) values of $ W $ is obtained
when positive (negative) temperatures
are used. In both cases, temperatures of large absolute $|\theta|$ value will
cause a sampling of the distribution close to its typical value,
while temperatures of small absolute value are used to access the
tails of the distribution.
Thus, performing the simulations for
a suitably chosen set of temperatures allows us to estimate $ P(W) $
over a large range, possibly on its full support.

The normalization constants $ Z(\theta) $ can be computed by
comparing histograms from neighboring temperatures. In their overlapping region,
relative normalization constants can be obtained by requiring that the rescaled
histograms must agree within error bars. This means the histograms are
``glued'' together. The range of covered $ W $ values can be extended
iteratively by choosing additional suitable temperatures and gluing the
resulting histograms to each other. For details, see Refs \cite{align2002,largest-2011}.

\section{Results\label{sec:results}}

We have performed simulations for complete graphs and connected ER
graphs with varying connectivity $c$ for different number $N$ of nodes
in the range $N\le 512$. For both ensembles,
the weights are uniformly distributed in $[0,1]$.
First, we consider the results for
the complete graphs; in the second subsection, the ER case.

\subsection{Complete graphs\label{sec:complete}}

\begin{figure}
    \includegraphics[width=0.99\columnwidth]{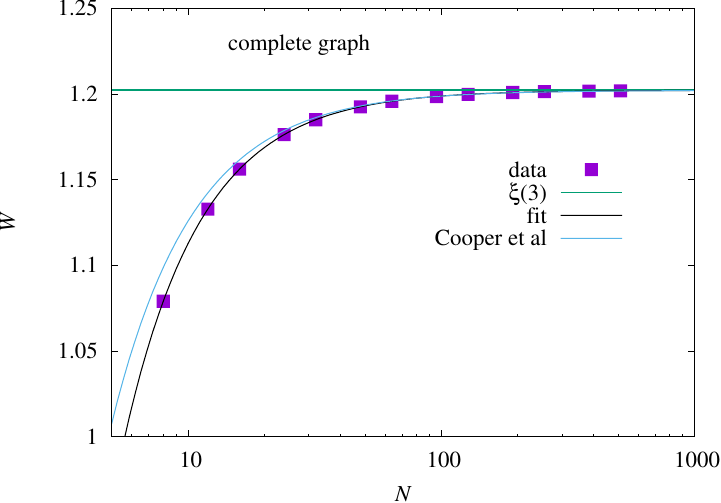}
    \caption{\label{fig:mean_N_full}
        Average spanning tree weight $W$ as a function of the number $N$ of nodes
        for the complete graph. The horizontal line indicates the limiting
        value obtained analytically, see Eq.~(\ref{eq:expectation}),
        while the upper solid curve prepresents
        Eq.\ (\ref{eq:power-lawB}). The lower solid
        curve indicates a fit to Eq.\ (\ref{eq:power-law}).}
\end{figure}

The average spanning-tree weight for the complete graph is shown as a function
of the graph size in Fig.~\ref{fig:mean_N_full}. A convergence to the
analytical result Eq.~(\ref{eq:expectation}) is well visible,
confirming our simulations. Note that there exist exact results
\cite{steele2002} for
small graphs $N\le 9$. The result,
$\frac{199462271}{184848378} \approx 1.07906$,
agrees to the five leading digits with our simulation
data. We have fitted
a power law function
\begin{equation}
    W_{\rm avg}(N)=W_{\rm avg}^{\infty}+ b N^{-c}\,,
    \label{eq:power-law}
\end{equation}
resulting in $W_{\rm avg}^{\infty}=1.20211(2)$, $b=-2.349(5)$,
and $c=1.418(1)$. As visible from the figure, the finite-size dependence of
the weight follows the power-law very well. On the other hand,
it has been predicted by Cooper et al. \cite{cooper2016} that the finite
corrections have the form
\begin{equation}
    W(N)=\zeta(3)+ c_1 N^{-1}+c_2 N^{-4/3}+\ldots\,,
    \label{eq:power-lawB}
\end{equation}
where $c_1=0.0384956\ldots$, $c_2=-1.7295$, and the $\ldots$
in the equation refers to even
faster decaying correction terms. We have included this function without a
free parameter in Fig.~\ref{fig:mean_N_full}. Here the agreement is worse,
showing that the additional corrections are significant.

\begin{figure}
    \includegraphics[width=0.99\columnwidth]{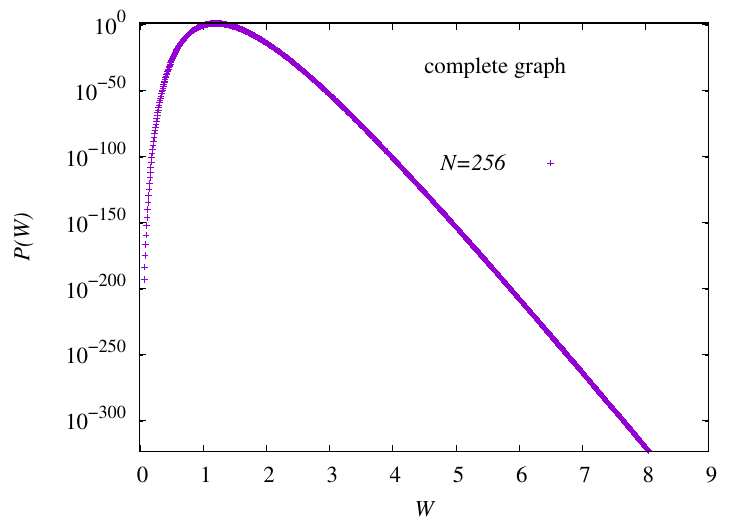}
    \caption{\label{fig:dist_complete}
        Distribution of spanning tree weight $W$ for the complete graph
        with $N=256$ nodes.}
\end{figure}

Next, we consider the full distribution $P(W)$.
In Fig.~\ref{fig:dist_complete} it is shown for
the complete graph with $N=256$ nodes. By using the
large-deviation approach, we are able to reach the tails of the
distributions to values of the probability density as small as $10^{-300}$.
Since $W>0$, we tried to fit the data to a generalized
extreme-value distribution with shape parameter $\xi>0$, but the fit did
not converge. Still,
the right tail resembles an exponential, but is slightly bent. A fit
of the right tail to a stretched exponential $e^{-\lambda(W-W_0)^{\beta}}$
(not shown) yields an exponent $\beta=1.22(2)$.

\begin{figure}
    \includegraphics[width=0.99\columnwidth]{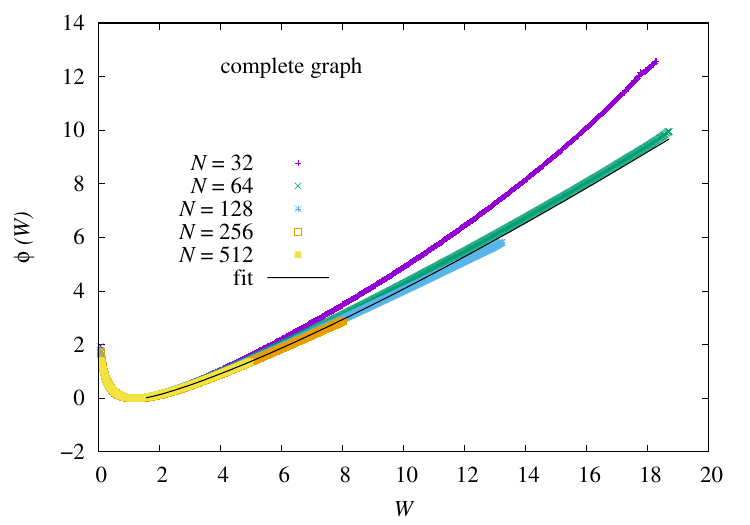}
    \caption{\label{fig:rate_function}
        Rate functions $\phi(W)$ for the complete graph for different
        values of the number $N$ of nodes. The line shows the result of a fit
        of the data for $N=256$ to a power law Eq.~(\ref{eq:power:law:rate}).}
\end{figure}

Next, we analyze the rate function Eq.~(\ref{eq:rate_function}), which is shown
for different number $N$ of nodes in Fig.~\ref{fig:rate_function}.
Note that for $N=512$ for the far tail of the distribution, the MCMC
simulations do not fully equilibrate. Thus, we cannot use this part of the data,
and we are restricted here to somewhat smaller values of $W$.
While for small values of $N$ some finite size effects are visible,
the curves, at least for values of $W$ up to about $6$, collapse onto each other
for $N\ge 128$. This means that the convergence of the empirical rate function
to a limiting rate function is compatible with the data. Still, we cannot say
much about the shape, but the compatibility of the tail with a stretched
exponential means that here the right part is compatible with a power law
\begin{equation}
    \tilde \phi(W)=l(W-W_0)^{\beta}
    \label{eq:power:law:rate}
\end{equation}
with the same exponent
$\beta=1.22(2)$ 
as mentioned above.

Now, we want to understand what makes rare graph realizations exhibit particular
large or small MST weights $W$. For the complete graph, the graph structure
is fixed, so it must be due to the specific edge weights, i.e.,
the $N(N-1)/2$ weights of each graph $G$. For this
purpose, we have stored, here for the largest considered size $N=512$,
for each statistically independent
graph $G=(V,E)$ encountered in the MCMC simulation, the mean
$W_{G}^{\text{mean}}= \frac 2 {N(N-1)} \sum_{\{i,j\}\in V} w_{i,j}$ weight.
Also, we have for each node determined the minimum
weight among all edges that are incident to the node.
The empirical average over all nodes we denote as $W_{G}^{\text{min}}$.
Since each node is connected to all other nodes by at least one
edge, $(N-1)W_{G}^{\text{min}}$ forms a lower bound and also a good
approximation of $W$. In a similar way,
since each spanning tree consists of $N-1$ edges, we have stored the weight
$W_{G}^{N-1}$ of the $(N-1)$'th largest weight in each graph $G$. This might
serve as a good upper approximation, but not a strict upper bound.
Finally, we also store the empirical
variance $W_{G}^{\text{var}}$ of the weights of each graph to see whether
all edge weights get concentrated around some values.

\begin{figure}
    \includegraphics[width=0.99\columnwidth]{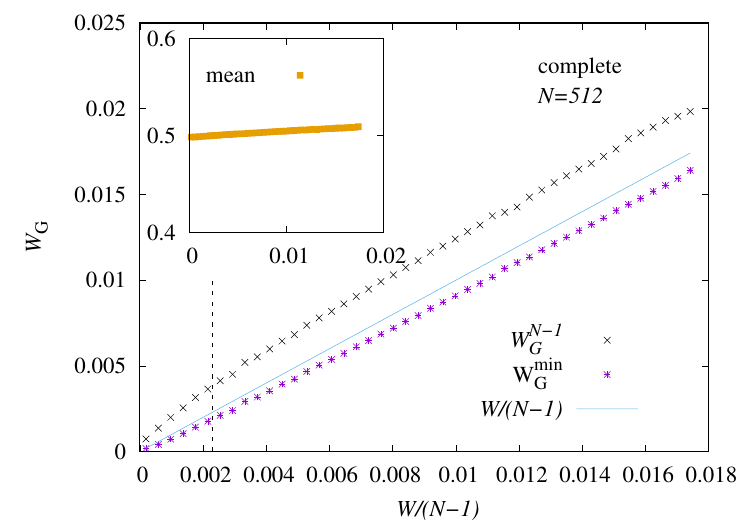}
    \caption{\label{fig:corr_complete}
        Average of $W_{G}^{\text{min}}$, $W_{G}^{N-1}$, and (inset) $W_{G}^{\text{mean}}$,
        conditioned to the value $W$ of the MST of $G$, respectively.
        The line indicates the diagonal $W_G=W/(N-1)$.
        The vertical line near $W/(N-1)=0.002$ indicates the typical
        edge weight of an MST, corresponding to $P(W)$ exhibiting a peak
        near $W=1$ in Fig.~\ref{fig:dist_complete} and $N=512$.
    }
\end{figure}

We analyze now these quantities as a function of the MST $W$ of each graph
$G$.
For this purpose, we show the averages of $W_{G}^{\text{min}}$, $W_{G}^{N-1}$,
$W_{G}^{\text{mean}}$ conditioned to $W=W(G)$ in Fig.~\ref{fig:corr_complete},
as a function, for better comparison, of $W/(N-1)$, i.e., the average
weight of an edge in the MST. One sees a strong correlation of $W$
with $W_{G}^{\text{min}}$ and $W_{G}^{N-1}$, and indeed $W/(N-1)$ lies
well between them.
The observed linear correlation coefficients are 0.9997 and 0.9971,
respectively.
A scatter plot (not shown) reveals
that the values of $W_{G}^{\text{min}}$ and $W_{G}^{N-1}$ are concentrated
near the typical values, conditioned to $W$.
For just a typical random graph, the MST weight is about 0.002,
indicated by the vertical line in the figure. The $N-1$ smallest edge weights
are below a weight of 0.004. This means that basically the smallest
weights contained in the graph actually contribute to the MST, although
the edges in the MST cannot be chosen independently.
On the other hand, each complete graph
contains many more edges, where most edges do not contribute to the MST.
As the inset of Fig.~\ref{fig:corr_complete} shows, a typical edge
weight is always near the average weight of 0.5, but one can indeed
observe
a clear but weaker correlation, with a linear correlation coefficient of
0.9946. The mean conditioned to large MST weights is 0.509, only slightly larger
than the unconditional mean of 0.5.
A similar weak effect can be seen for the empirical
variance (not shown as a figure), where we observed a
clear linear correlation coefficient of -0.9637, which results
from a small decay of the variance
from 0.0839 for $W/(N-1)=0.0005$ to a variance of 0.0805 for $W/(N-1)=0.018$.

Note that such a large range for the values of $W/(N-1)\in [0,0.018]$
can only be accessed with
a large-deviation approach. With a direct sampling of, say, $10^6$
samples, one would be able to observe values and therefore measure
correlations for about
$W/(N-1)\in [0.002,0.003]$.

\subsection{Erd\H{o}s-R\'enyi graphs}

\begin{figure}
    \includegraphics[width=0.99\columnwidth]{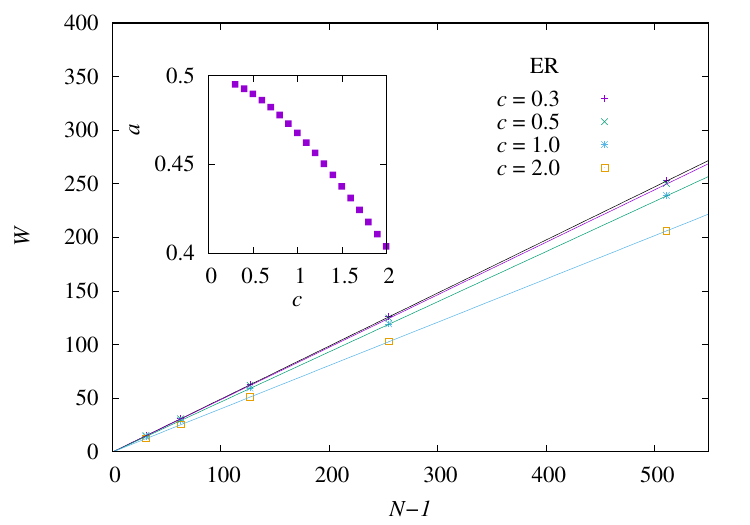}
    \caption{\label{fig:mean_N_ER}
        Average spanning tree weight $W$ as a function of the number $N$ of nodes
        for the connected ER ensemble. The lines indicate fits to a linear
        function with slope $a$. The inset shows the slope $a$ as a function
        of the connectivity $c$.}
\end{figure}

For the connected ER ensemble, the behavior can be richer, since not
only the edge weights but also the graph structure varies.
We first analyse the mean minimum spanning tree weight, which
we denote by $W$ as well for simplicity, as a function of
the number $N$ of nodes. Since the
connected ER ensemble is sparse, the number of available edges to
construct an MST is much smaller, of the order $O(N)$.
Thus, doubling the size of the graph will basically double the weight
of the MST; hence, $W$, being the sum of $N-1$ weights,
should scale linearly with $N-1$.
This is well visible in Fig.~\ref{fig:mean_N_ER}. We have fitted the data
for each value of $c$ to a linear function $W=a(N-1)$.
For small values
of $c$, the original ER graphs contain $cN/2$ edges, which is smaller than
$N-1$, the number of edges required to form a spanning tree. Hence,
the connected ER ensemble will usually contain $N-1$ edges or a few more,
despite the small value of $c$. Thus,
almost all edges are contained in the spanning tree, which means that,
because the average edge weight is $1/2$,
$W$ will be about $(N-1)/2$, i.e., $a=1/2$.
As shown in the inset of
Fig.~\ref{fig:mean_N_ER}, the obtained values of $a$ are indeed close
to $1/2$ for small values of $c$. Note that the value $c=1$,
which is the percolation transition point of the standard ER ensemble,
does not show any special behavior here.

\begin{figure}
    \includegraphics[width=0.99\columnwidth]{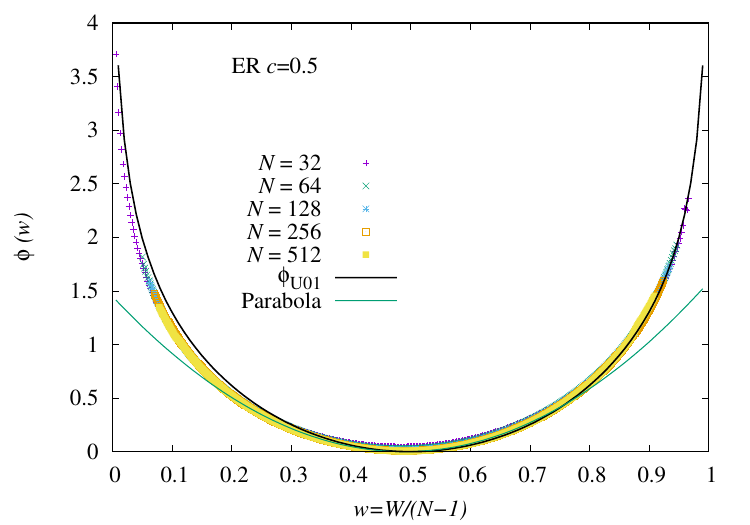}
    \caption{\label{fig:rate_ER05}
        Rate functions $\phi(W)$ for the connected ER graph with $c=0.5$
        for different
        values of the number $N$ of nodes. The black line shows
        the rate function $\phi_{U01}$ for the average of $N-1$
        iid random numbers distributed uniformly in $[0,1]$.
        The green line shows the result of a fit
        of the data for $N=512$ in the high-probability
        region to a parabola Eq.~(\ref{eq:parabola:rate}).}
\end{figure}

Next, to investigate the actual distributions,
we consider the rate functions $\phi(w)$ with $w=W/(N-1)$
being the spanning-tree
weight per edge. The result for $c=0.5$ is shown in Fig.~\ref{fig:rate_ER05}.
The empirical rate functions for different sizes collapse
almost onto each other, so a convergence to a limiting rate function
in the limit of large graph sizes can be expected, and the large-deviation
property is fulfilled.

As mentioned above, the restriction to connected graphs means we expect
that here there are basically graphs with $N-1$ edges for $c=0.5$ such that
almost all edges contribute to the spanning tree. Thus, the spanning tree
weight is very simply almost given by
the sum of $N-1$ random numbers that are iid in $[0,1]$, i.e., $W/(N-1)$
is almost the average weight of these numbers.

For a first simple comparison, we consider the region of larger
probabilities.
From the central limit theorem, we therefore expect that
the typical part of the distribution is Gaussian $\exp(-(w-w_p)^2/(2\sigma^2))$,
centered about
the mean value 0.5 and with variance $\sigma^2=1/(12 (N-1))$, with
1/12 being the variance of the uniform
distribution in $[0,1]$. This is reflected visibly
in the shape of the rate function. Indeed, a fit to a parabola
\begin{equation}
    \tilde \phi(w)=l(w-w_0)^2
    \label{eq:parabola:rate}
\end{equation}
in the range $w\in[0.3,0.7]$ yields a good fit; see Fig.~\ref{fig:rate_ER05}.
Here we find $w_0=0.489(1)$ and $l=6.08$. The differences
to the expected values $w_0=0.5$ and $l=12/2$, respectively,
are likely due to the finite size of the system.
E.g., the variance behaves like $1/12 (N-1)$, but for the rate function
we divide by $N$ instead of $N-1$.
The result of the fit confirms the expectations
about the shape of minimum spanning trees in this low-connectivity
region.

To compare also the tails, the rate function of the average
of $N-1$ iid $U(0,1)$
random numbers can be calculated. This is the correct model
for a tree, where all edges contribute to the spanning tree;
see above. To obtain
the rate function of a random variable (RV) $S_N$ using the
G\"{a}rtner-Ellis Theorem \cite{touchette2009},
we need the scaled cumulant generating function
$\lambda(k) = \lim_{N\to\infty} \ln \frac 1 N E[e^{N k S_N}]$,
where the expectation $E[\cdot]$
is with respect to the $N$ random variables $\{X_i \}$
contributing to $S_N$.
For the average $S_N= \frac 1 N \sum_i X_i$, where all RVs $X_i$
are the same $X$,
we have
$\lambda(k)=\lim_{N\to\infty} \ln \frac 1 N E[e^{k \sum_i X_i}]$
$= \ln E[e^{k X}]_X$, where $E[\cdot]_X$ is the average of
the random RV $X$. For the uniform distribution $U(0,1)$, this yields
$\lambda(k)=\ln \int_0^1 dx\, e^{k x}$ =$\ln((e^k-1)/k)$. By using the
G\"{a}rtner-Ellis Theorem, one has $\phi(s)=\sup_{k} (ks-\lambda(k))$.
We have performed this maximization numerically for many values of
$s\in[0,1]$; the result, which we denote as $\phi_{U01}$, is also shown in
Fig.~\ref{fig:rate_ER05}. A very good agreement with the numerical
data is visible; the small differences are likely due to the
fact that also graphs with more than $N-1$ edges contribute,
in particular for small values of $w$.

We have also considered $c=0.7$ and $c=1$ in the (about)
non-percolating region; here the behavior is very
similar.

\begin{figure}
    \includegraphics[width=0.99\columnwidth]{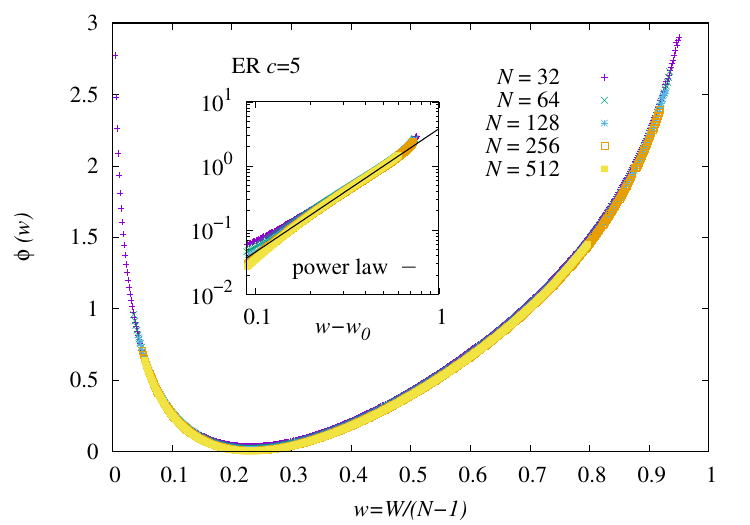}
    \caption{\label{fig:rate_ER5}
        Rate functions $\phi(W)$ for the connected ER graph with $c=5$
        for different
        values of the number $N$ of nodes. In the inset,
        the same data is shown with a double log scale. The line shows
        the result of a fit
        of the data for $N=512$ to a power law Eq.~(\ref{eq:power:law:rate}).}
\end{figure}

For larger values of $c$, the shape of the distribution changes.
In Fig.~\ref{fig:rate_ER5}, we show the rate functions for the largest
value of $c=5$ we have considered. The function is clearly not symmetric.
For larger values of $N$, no significant finite-size dependence
is visible, indicating that a convergence $N\to\infty$ exists and
therefore the large-deviation principle holds also here. In the intermediate
regime, $w\in [0.3,0.7]$, we have fitted a power law
Eq.~(\ref{eq:power:law:rate}) yielding an offset $w_0= 0.19$
and an exponent $\beta=1.92(2)$, which is close to but significantly
different from $2$. This matches the fact that for small values of $c$,
a parabolic shape $\beta=2$ was found, while for the complete graphs,
corresponding to $c\to \infty$, in Sec.~\ref{sec:complete},
a much smaller exponent was obtained.

Note that for larger values of $w$, the data bends up; thus, a growth with
a larger exponent is possible. Such change of the behavior
of rate functions, sometimes termed ``phase transitions'', has
been observed also for the rate functions of other random
systems like those described by the Kardar-Parisi-Zhang equation \cite{kpz2018}.

\begin{figure}
    \includegraphics[width=0.99\columnwidth]{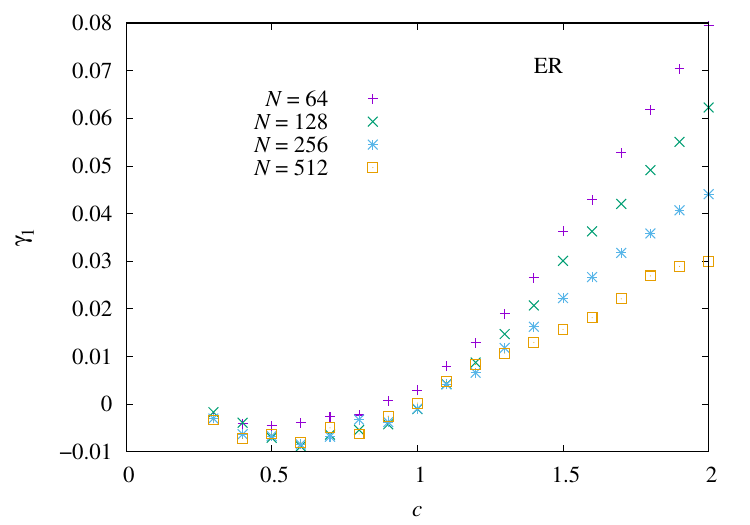}
    \caption{\label{fig:skewness_c_ER}
        Skewness $\gamma_1$ as a function of the connectivity parameter $c$
        for the connected ER ensemble, for different system sizes $N$.}
\end{figure}

To investigate more how the shape of the distribution changes when varying
the parameter $c$, we analyse the \emph{skewness}
\begin{equation}
    \gamma_1 = \frac{\langle (W-\langle W \rangle)^3\rangle }{\sigma^3}\,,
\end{equation}
where $\sigma^2$ is the variance $\langle (W -\langle W \rangle)^2\rangle$.
The result is shown for different number $N$ of nodes in
Fig.~\ref{fig:skewness_c_ER}. For values $c\le 1$, the skewness is very close
to zero, indicating a very symmetric distribution. Only from $c=1$
onwards a gradual increase is visible. Thus, for values of $c$
that correspond to the percolating phase in the standard ER ensemble,
the MST weight of the connected ER ensemble exhibits a significant asymmetric
distribution.

\begin{figure}
    \includegraphics[width=0.99\columnwidth]{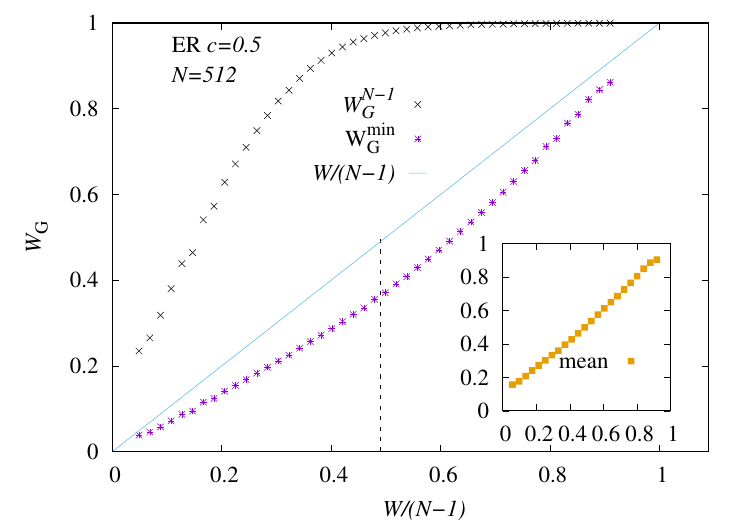}
    \caption{\label{fig:corr_ER05}
        Average of $W_{G}^{\text{min}}$, $W_{G}^{N-1}$, and (inset) $W_{G}^{\text{mean}}$,
        conditioned to the value $W$ of the MST of $G$, respectively,
        for ER random graphs with $c=0.5$.
        The line indicates the diagonal $W_G=W/(N-1)$.
        The vertical line near $W/(N-1)=0.49$ indicates the typical
        edge weight in an MST for this value of $c$.
    }
\end{figure}

Finally, we investigate how the structure of the graphs influences the
MST weight. Similar to the results for the complete graph, we show
in Fig.~\ref{fig:corr_ER05} for the case $c=0.5$ the
average graph weights $W_{G}^{\text{min}}$, $W_{G}^{N-1}$,
and $W_{G}^{\text{mean}}$
conditioned to a value of the weight $W/(N-1)$ per MST edge.
Again we see that the average edge weight
is between $W_{G}^{\text{min}}$ and $W_{G}^{N-1}$. Compared to the complete graph,
see Fig.~\ref{fig:corr_complete},
the upper value $W_{G}^{N-1}$ is farther away from $W/(N-1)$. The reason is
that for the connected ER graphs of such a small value of $c$, as discussed
above, almost all edges contribute to the MST. Thus, $W_{G}^{N-1}$
will be close to the upper limit of $w=1$ for typical graphs and
for atypical graphs with larger MST weights, as visible in
Fig.~\ref{fig:corr_ER05}. Only for atypical small values of $W$,
$W_{G}^{N-1}$ is likely smaller. Also, since almost all edges contribute
to the MST, the average weight $W/(N-1)$
of the MST is very  similar to the
average edge weight $W_{G}^{\text{mean}}$ in the graph, as visible in the inset.

\begin{figure}
    \includegraphics[width=0.99\columnwidth]{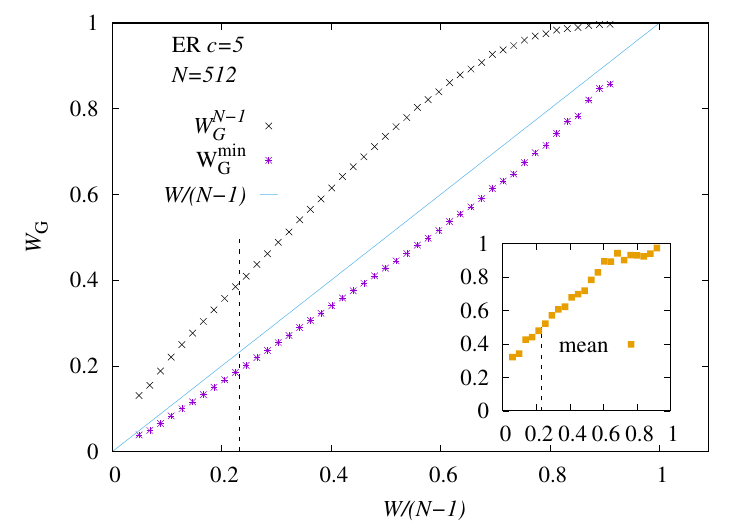}
    \caption{\label{fig:corr_ER50}
        Average of $W_{G}^{\text{min}}$, $W_{G}^{N-1}$, and (inset) $W_{G}^{\text{mean}}$,
        conditioned to the value $W$ of the MST of $G$, respectively,
        for ER random graphs with $c=5$.
        The line indicates the diagonal $W_G=W/(N-1)$.
        The vertical line near $W/(N-1)=0.23$ indicates the typical
        edge weight in an MST for this value of $c$.
    }
\end{figure}

The corresponding results for $c=5$ are displayed in Fig.~\ref{fig:corr_ER50}.
Since more edges are present in the graph, but still the graph is sparse,
the results look somehow in between the results for $c=0.5$ and for the
complete graph. Again, here the sparseness leads to a high influence
of the mean edge weight $W_{G}^{\text{mean}}$ of $G$ on the MST weight $W$.
We do not display results for the other values of $c$ that we have studied
in the same way,
because they are intermediate between the two cases presented so far.

\begin{figure}
    \includegraphics[width=0.99\columnwidth]{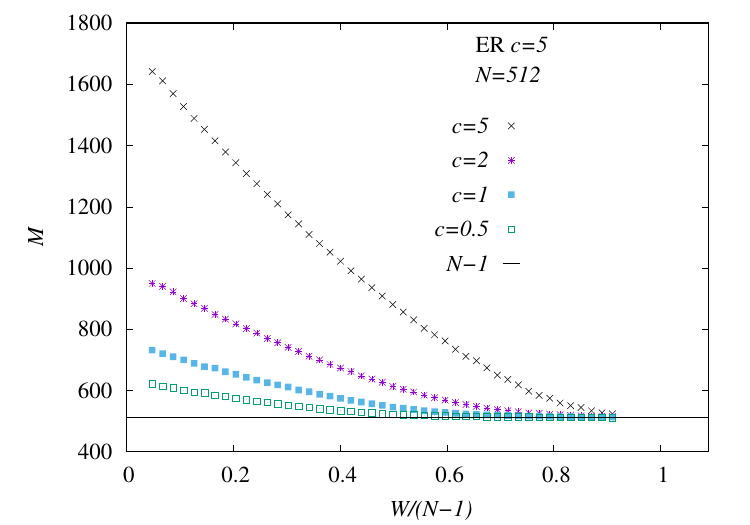}
    \caption{\label{fig:M_W_ER}
        Average of the number $M$ of edges
        conditioned to the value $W$ of the MST of $G$, shown as a function
        of $W/(N-1)$ for $N=512$ and several values of $c$.
        The horizontal line indicates the minimum number $N-1$ of
        edges needed to create a connected graph.
    }
\end{figure}

The fluctuations of the edge weights of the graph are not the only reason for
particular small or large MST weights.
In the ER ensemble, also the number of edges fluctuates. For any value of $c$,
although the typical number of edges is $cN/2$, there will be much denser
and much sparser graphs. In general, the more edges are available, the
smaller the MST weight will be, because there are more weights to minimize
over. This is visible in Fig.~\ref{fig:M_W_ER}, where the average number
$M$ of edges conditioned to the MST edge weight $W/(N-1)$ per edge
is shown for $N=512$ and several values of $c$.
For large values of $W$, the number of edges is very small
and converges to the lower limit of $N-1$, since the graphs have
to be connected. The dependency becomes more pronounced with increasing
value of $c$. This is reasonable because for smaller values of $c$
also typical connected graphs contain only few egdes.
Still, since the value $W/(N-1)$ is larger than
the mean weight 0.5, these graphs are not only rare with respect to
their structure, but also rare with respect to the edge weights,
as discussed above. 
For small values of $W/(N-1)$, the number of
edges is conversely atypically larger, so the minimization can lead
to lower MST weights.

In summary, both atypiacl graph structure and atypical
edge weights come together for weighted connected ER graphs which exhibit
atypical MST weights.

\section{Discussion\label{sec:discussion}}

We have studied the distribution of the weight of minimum-spanning trees
for complete graphs as well as for connected ER random graphs
with connectivity $c$.
With respect to simple properties as the expectation value, our results
are compatible with previous analytical studies. By applying large-deviation
approaches, we are able to obtain the distributions even in the tails,
down to probability densities as small as $10^{-300}$. For both ensembles,
the results indicate that the large-deviation property is fulfilled, i.e.,
away from the typical values the probabilities become exponentially small
with exponent linearly decreasing in $N$.
For the complete graph, the behavior of the rate function of the right tails
is compatible with a power-law behavior. Since for complete graphs,
the graph structure is fixed, rare MST weights $W$ are created by
rare assignments of edge weights. But only the lowest edge weights
have to be smaller or larger than typical to determine $W$,
the overall mean weight in a graph,
is only weakly affected.

For the ER case, for small values of $c$, the distributions converge
in the center to Gaussians, which makes sense
as many independent weights contribute. Also the tails of the
distribution are described well by a rate function describing the
sum of $N-1$ independently drawn numbers in $[0,1]$.
Beyond the value $c=1$, which is the percolation threshold for the
standard ER ensemble, the distributions become asymmetric
and non-trivial tails develop. Here, the values of edge weights
and the actual number of edges influence
the MST weight and determine whether it is typical, atypically large or
atypically small.

Clearly, one could analyse more properties, e.g.,
whether the variance of the node degrees also play a, likely minor, role.
For future studies, one also could consider other ensembles like scale-free
graphs, finite-dimensional disordered lattices, or general geometric graphs
embedded in finite-dimensional spaces.

Also, it would be of particular interest if our work motivates other
researchers to study the distribution of weights by analytical approaches,
e.g., to obtain the rate functions and to investigate whether there
is a ``critical'' value $w_c$ where the rate function changes its behavior
for the ER case.

\begin{acknowledgments}

    MS thanks Abolfazl Ramezanpour for helpful discussions and assistance with debugging the code.
    The simulations were performed at the
    HPC cluster ROSA, located at the University of Oldenburg (Germany) and
    funded by the DFG through its Major Research Instrumentation Program
    (INST 184/225-1 FUGG) and the Ministry of Science and Culture (MWK) of the
    Lower Saxony State.

\end{acknowledgments}

\bibliography{refs.bib}

\begin{thebibliography}{45}%
\makeatletter
\providecommand \@ifxundefined [1]{%
 \@ifx{#1\undefined}
}%
\providecommand \@ifnum [1]{%
 \ifnum #1\expandafter \@firstoftwo
 \else \expandafter \@secondoftwo
 \fi
}%
\providecommand \@ifx [1]{%
 \ifx #1\expandafter \@firstoftwo
 \else \expandafter \@secondoftwo
 \fi
}%
\providecommand \natexlab [1]{#1}%
\providecommand \enquote  [1]{``#1''}%
\providecommand \bibnamefont  [1]{#1}%
\providecommand \bibfnamefont [1]{#1}%
\providecommand \citenamefont [1]{#1}%
\providecommand \href@noop [0]{\@secondoftwo}%
\providecommand \href [0]{\begingroup \@sanitize@url \@href}%
\providecommand \@href[1]{\@@startlink{#1}\@@href}%
\providecommand \@@href[1]{\endgroup#1\@@endlink}%
\providecommand \@sanitize@url [0]{\catcode `\\12\catcode `\$12\catcode
  `\&12\catcode `\#12\catcode `\^12\catcode `\_12\catcode `\%12\relax}%
\providecommand \@@startlink[1]{}%
\providecommand \@@endlink[0]{}%
\providecommand \url  [0]{\begingroup\@sanitize@url \@url }%
\providecommand \@url [1]{\endgroup\@href {#1}{\urlprefix }}%
\providecommand \urlprefix  [0]{URL }%
\providecommand \Eprint [0]{\href }%
\providecommand \doibase [0]{http://dx.doi.org/}%
\providecommand \selectlanguage [0]{\@gobble}%
\providecommand \bibinfo  [0]{\@secondoftwo}%
\providecommand \bibfield  [0]{\@secondoftwo}%
\providecommand \translation [1]{[#1]}%
\providecommand \BibitemOpen [0]{}%
\providecommand \bibitemStop [0]{}%
\providecommand \bibitemNoStop [0]{.\EOS\space}%
\providecommand \EOS [0]{\spacefactor3000\relax}%
\providecommand \BibitemShut  [1]{\csname bibitem#1\endcsname}%
\let\auto@bib@innerbib\@empty
\bibitem [{\citenamefont {Albert}\ and\ \citenamefont
  {Barab\'asi}(2002)}]{albert2002review}%
  \BibitemOpen
  \bibfield  {author} {\bibinfo {author} {\bibfnamefont {R.}~\bibnamefont
  {Albert}}\ and\ \bibinfo {author} {\bibfnamefont {A.-L.}\ \bibnamefont
  {Barab\'asi}},\ }\href@noop {} {\bibfield  {journal} {\bibinfo  {journal}
  {Rev. Mod. Phys.}\ }\textbf {\bibinfo {volume} {74}},\ \bibinfo {pages} {47}
  (\bibinfo {year} {2002})}\BibitemShut {NoStop}%
\bibitem [{\citenamefont {Newman}(2003)}]{newman2003review}%
  \BibitemOpen
  \bibfield  {author} {\bibinfo {author} {\bibfnamefont {M.~E.~J.}\
  \bibnamefont {Newman}},\ }\href@noop {} {\bibfield  {journal} {\bibinfo
  {journal} {SIAM Review}\ }\textbf {\bibinfo {volume} {45}},\ \bibinfo {pages}
  {167} (\bibinfo {year} {2003})}\BibitemShut {NoStop}%
\bibitem [{\citenamefont {Boccaletti}\ \emph {et~al.}(2006)\citenamefont
  {Boccaletti}, \citenamefont {Latora}, \citenamefont {Moreno}, \citenamefont
  {Chavez},\ and\ \citenamefont {Hwang}}]{boccaletti2006}%
  \BibitemOpen
  \bibfield  {author} {\bibinfo {author} {\bibfnamefont {S.}~\bibnamefont
  {Boccaletti}}, \bibinfo {author} {\bibfnamefont {V.}~\bibnamefont {Latora}},
  \bibinfo {author} {\bibfnamefont {Y.}~\bibnamefont {Moreno}}, \bibinfo
  {author} {\bibfnamefont {M.}~\bibnamefont {Chavez}}, \ and\ \bibinfo {author}
  {\bibfnamefont {D.~U.}\ \bibnamefont {Hwang}},\ }\href@noop {} {\bibfield
  {journal} {\bibinfo  {journal} {Phys.\ Rep.}\ }\textbf {\bibinfo {volume}
  {424}},\ \bibinfo {pages} {175} (\bibinfo {year} {2006})}\BibitemShut
  {NoStop}%
\bibitem [{\citenamefont {Dorogovtsev}\ and\ \citenamefont
  {Mendes}(2006)}]{dorogovtsev_book2006}%
  \BibitemOpen
  \bibfield  {author} {\bibinfo {author} {\bibfnamefont {S.~N.}\ \bibnamefont
  {Dorogovtsev}}\ and\ \bibinfo {author} {\bibfnamefont {J.~F.~F.}\
  \bibnamefont {Mendes}},\ }\href@noop {} {\emph {\bibinfo {title} {Evolution
  of networks: from biological nets to the Internet and WWW}}}\ (\bibinfo
  {publisher} {Oxford Univ.\ Press},\ \bibinfo {year} {2006})\BibitemShut
  {NoStop}%
\bibitem [{\citenamefont {Newman}(2010)}]{newman_book2010}%
  \BibitemOpen
  \bibfield  {author} {\bibinfo {author} {\bibfnamefont {M.}~\bibnamefont
  {Newman}},\ }\href@noop {} {\emph {\bibinfo {title} {Networks: an
  Introduction}}}\ (\bibinfo  {publisher} {Oxford University Press},\ \bibinfo
  {year} {2010})\BibitemShut {NoStop}%
\bibitem [{\citenamefont {Barrat}\ \emph {et~al.}(2012)\citenamefont {Barrat},
  \citenamefont {Barth\'elemy},\ and\ \citenamefont
  {Vespignani}}]{barrat_book2012}%
  \BibitemOpen
  \bibfield  {author} {\bibinfo {author} {\bibfnamefont {A.}~\bibnamefont
  {Barrat}}, \bibinfo {author} {\bibfnamefont {M.}~\bibnamefont
  {Barth\'elemy}}, \ and\ \bibinfo {author} {\bibfnamefont {A.}~\bibnamefont
  {Vespignani}},\ }\href@noop {} {\emph {\bibinfo {title} {Dynamical Processes
  on Complex Networks}}}\ (\bibinfo  {publisher} {Cambridge University Press},\
  \bibinfo {year} {2012})\BibitemShut {NoStop}%
\bibitem [{\citenamefont {Graham}\ and\ \citenamefont
  {Hell}(1985)}]{graham1985}%
  \BibitemOpen
  \bibfield  {author} {\bibinfo {author} {\bibfnamefont {R.~L.}\ \bibnamefont
  {Graham}}\ and\ \bibinfo {author} {\bibfnamefont {P.}~\bibnamefont {Hell}},\
  }\href@noop {} {\bibfield  {journal} {\bibinfo  {journal} {Ann. Hist.
  Comput.}\ }\textbf {\bibinfo {volume} {7}},\ \bibinfo {pages} {43} (\bibinfo
  {year} {1985})}\BibitemShut {NoStop}%
\bibitem [{\citenamefont {Prim}(1957)}]{prim1957}%
  \BibitemOpen
  \bibfield  {author} {\bibinfo {author} {\bibfnamefont {R.~C.}\ \bibnamefont
  {Prim}},\ }\href {\doibase 10.1002/j.1538-7305.1957.tb01515.x} {\bibfield
  {journal} {\bibinfo  {journal} {The Bell System Technical Journal}\ }\textbf
  {\bibinfo {volume} {36}},\ \bibinfo {pages} {1389} (\bibinfo {year}
  {1957})}\BibitemShut {NoStop}%
\bibitem [{\citenamefont {Cavalli-Sforza}\ and\ \citenamefont
  {Edwards}(1967)}]{cavalli-sforza1967}%
  \BibitemOpen
  \bibfield  {author} {\bibinfo {author} {\bibfnamefont {L.~L.}\ \bibnamefont
  {Cavalli-Sforza}}\ and\ \bibinfo {author} {\bibfnamefont {A.~W.~F.}\
  \bibnamefont {Edwards}},\ }\href@noop {} {\bibfield  {journal} {\bibinfo
  {journal} {Evolution}\ }\textbf {\bibinfo {volume} {21}},\ \bibinfo {pages}
  {550} (\bibinfo {year} {1967})}\BibitemShut {NoStop}%
\bibitem [{\citenamefont {Gower}\ and\ \citenamefont {Ross}(1969)}]{gower1969}%
  \BibitemOpen
  \bibfield  {author} {\bibinfo {author} {\bibfnamefont {J.~C.}\ \bibnamefont
  {Gower}}\ and\ \bibinfo {author} {\bibfnamefont {G.~J.~S.}\ \bibnamefont
  {Ross}},\ }\href@noop {} {\bibfield  {journal} {\bibinfo  {journal} {J. Royal
  Stat. Soc. Ser. C}\ }\textbf {\bibinfo {volume} {18}},\ \bibinfo {pages} {54}
  (\bibinfo {year} {1969})}\BibitemShut {NoStop}%
\bibitem [{\citenamefont {Wu}\ \emph {et~al.}(2013)\citenamefont {Wu},
  \citenamefont {Li}, \citenamefont {Jiao}, \citenamefont {Wang},\ and\
  \citenamefont {Sun}}]{wu2013}%
  \BibitemOpen
  \bibfield  {author} {\bibinfo {author} {\bibfnamefont {J.}~\bibnamefont
  {Wu}}, \bibinfo {author} {\bibfnamefont {X.}~\bibnamefont {Li}}, \bibinfo
  {author} {\bibfnamefont {L.}~\bibnamefont {Jiao}}, \bibinfo {author}
  {\bibfnamefont {X.}~\bibnamefont {Wang}}, \ and\ \bibinfo {author}
  {\bibfnamefont {B.}~\bibnamefont {Sun}},\ }\href {\doibase
  10.1016/j.physa.2013.01.015} {\bibfield  {journal} {\bibinfo  {journal}
  {Physica A}\ }\textbf {\bibinfo {volume} {392}},\ \bibinfo {pages} {2265}
  (\bibinfo {year} {2013})}\BibitemShut {NoStop}%
\bibitem [{\citenamefont {Osteen}\ and\ \citenamefont
  {Lin}(1974)}]{osteen1974}%
  \BibitemOpen
  \bibfield  {author} {\bibinfo {author} {\bibfnamefont {R.~E.}\ \bibnamefont
  {Osteen}}\ and\ \bibinfo {author} {\bibfnamefont {P.~P.}\ \bibnamefont
  {Lin}},\ }\href {\doibase 10.1137/0203003} {\bibfield  {journal} {\bibinfo
  {journal} {SIAM J. Comput.}\ }\textbf {\bibinfo {volume} {3}},\ \bibinfo
  {pages} {23} (\bibinfo {year} {1974})}\BibitemShut {NoStop}%
\bibitem [{\citenamefont {Zhang}\ \emph {et~al.}(2012)\citenamefont {Zhang},
  \citenamefont {Wu},\ and\ \citenamefont {Lin}}]{zhang2012}%
  \BibitemOpen
  \bibfield  {author} {\bibinfo {author} {\bibfnamefont {Z.}~\bibnamefont
  {Zhang}}, \bibinfo {author} {\bibfnamefont {B.}~\bibnamefont {Wu}}, \ and\
  \bibinfo {author} {\bibfnamefont {Y.}~\bibnamefont {Lin}},\ }\href {\doibase
  10.1016/j.physa.2012.01.039} {\bibfield  {journal} {\bibinfo  {journal}
  {Physica A}\ }\textbf {\bibinfo {volume} {391}},\ \bibinfo {pages} {3342}
  (\bibinfo {year} {2012})}\BibitemShut {NoStop}%
\bibitem [{\citenamefont {Li}\ and\ \citenamefont {Yan}(2019)}]{li2019}%
  \BibitemOpen
  \bibfield  {author} {\bibinfo {author} {\bibfnamefont {T.}~\bibnamefont
  {Li}}\ and\ \bibinfo {author} {\bibfnamefont {W.}~\bibnamefont {Yan}},\
  }\href {\doibase 10.1016/j.physa.2019.04.113} {\bibfield  {journal} {\bibinfo
   {journal} {Physica A}\ }\textbf {\bibinfo {volume} {536}},\ \bibinfo {pages}
  {120877} (\bibinfo {year} {2019})}\BibitemShut {NoStop}%
\bibitem [{\citenamefont {Dussert}\ \emph {et~al.}(1986)\citenamefont
  {Dussert}, \citenamefont {Rasigni}, \citenamefont {Rasigni}, \citenamefont
  {Palmari},\ and\ \citenamefont {Llebaria}}]{dussert1986}%
  \BibitemOpen
  \bibfield  {author} {\bibinfo {author} {\bibfnamefont {C.}~\bibnamefont
  {Dussert}}, \bibinfo {author} {\bibfnamefont {G.}~\bibnamefont {Rasigni}},
  \bibinfo {author} {\bibfnamefont {M.}~\bibnamefont {Rasigni}}, \bibinfo
  {author} {\bibfnamefont {J.}~\bibnamefont {Palmari}}, \ and\ \bibinfo
  {author} {\bibfnamefont {A.}~\bibnamefont {Llebaria}},\ }\href {\doibase
  10.1103/PhysRevB.34.3528} {\bibfield  {journal} {\bibinfo  {journal} {Phys.
  Rev. B}\ }\textbf {\bibinfo {volume} {34}},\ \bibinfo {pages} {3528}
  (\bibinfo {year} {1986})}\BibitemShut {NoStop}%
\bibitem [{\citenamefont {Ktitarev}\ \emph {et~al.}(2000)\citenamefont
  {Ktitarev}, \citenamefont {L\"ubeck}, \citenamefont {Grassberger},\ and\
  \citenamefont {B.~Priezzhev}}]{ktitarev2000}%
  \BibitemOpen
  \bibfield  {author} {\bibinfo {author} {\bibfnamefont {D.~V.}\ \bibnamefont
  {Ktitarev}}, \bibinfo {author} {\bibfnamefont {S.}~\bibnamefont {L\"ubeck}},
  \bibinfo {author} {\bibfnamefont {P.}~\bibnamefont {Grassberger}}, \ and\
  \bibinfo {author} {\bibfnamefont {V.}~\bibnamefont {B.~Priezzhev}},\ }\href
  {\doibase 10.1103/PhysRevE.61.81} {\bibfield  {journal} {\bibinfo  {journal}
  {Phys. Rev. E}\ }\textbf {\bibinfo {volume} {61}},\ \bibinfo {pages} {81}
  (\bibinfo {year} {2000})}\BibitemShut {NoStop}%
\bibitem [{\citenamefont {Wu}\ \emph {et~al.}(2006)\citenamefont {Wu},
  \citenamefont {Braunstein}, \citenamefont {Havlin},\ and\ \citenamefont
  {Stanley}}]{wu2006}%
  \BibitemOpen
  \bibfield  {author} {\bibinfo {author} {\bibfnamefont {Z.}~\bibnamefont
  {Wu}}, \bibinfo {author} {\bibfnamefont {L.~A.}\ \bibnamefont {Braunstein}},
  \bibinfo {author} {\bibfnamefont {S.}~\bibnamefont {Havlin}}, \ and\ \bibinfo
  {author} {\bibfnamefont {H.~E.}\ \bibnamefont {Stanley}},\ }\href {\doibase
  10.1103/PhysRevLett.96.148702} {\bibfield  {journal} {\bibinfo  {journal}
  {Phys. Rev. Lett.}\ }\textbf {\bibinfo {volume} {96}},\ \bibinfo {pages}
  {148702} (\bibinfo {year} {2006})}\BibitemShut {NoStop}%
\bibitem [{\citenamefont {Kim}\ \emph {et~al.}(2004)\citenamefont {Kim},
  \citenamefont {Noh},\ and\ \citenamefont {Jeong}}]{kim2004}%
  \BibitemOpen
  \bibfield  {author} {\bibinfo {author} {\bibfnamefont {D.-H.}\ \bibnamefont
  {Kim}}, \bibinfo {author} {\bibfnamefont {J.~D.}\ \bibnamefont {Noh}}, \ and\
  \bibinfo {author} {\bibfnamefont {H.}~\bibnamefont {Jeong}},\ }\href
  {\doibase 10.1103/PhysRevE.70.046126} {\bibfield  {journal} {\bibinfo
  {journal} {Phys. Rev. E}\ }\textbf {\bibinfo {volume} {70}},\ \bibinfo
  {pages} {046126} (\bibinfo {year} {2004})}\BibitemShut {NoStop}%
\bibitem [{\citenamefont {Bonanno}\ \emph {et~al.}(2003)\citenamefont
  {Bonanno}, \citenamefont {Caldarelli}, \citenamefont {Lillo},\ and\
  \citenamefont {Mantegna}}]{bonanno2003}%
  \BibitemOpen
  \bibfield  {author} {\bibinfo {author} {\bibfnamefont {G.}~\bibnamefont
  {Bonanno}}, \bibinfo {author} {\bibfnamefont {G.}~\bibnamefont {Caldarelli}},
  \bibinfo {author} {\bibfnamefont {F.}~\bibnamefont {Lillo}}, \ and\ \bibinfo
  {author} {\bibfnamefont {R.~N.}\ \bibnamefont {Mantegna}},\ }\href {\doibase
  10.1103/PhysRevE.68.046130} {\bibfield  {journal} {\bibinfo  {journal} {Phys.
  Rev. E}\ }\textbf {\bibinfo {volume} {68}},\ \bibinfo {pages} {046130}
  (\bibinfo {year} {2003})}\BibitemShut {NoStop}%
\bibitem [{\citenamefont {Onnela}\ \emph {et~al.}(2003)\citenamefont {Onnela},
  \citenamefont {Chakraborti}, \citenamefont {Kaski}, \citenamefont
  {Kert\'esz},\ and\ \citenamefont {Kanto}}]{onnela2003}%
  \BibitemOpen
  \bibfield  {author} {\bibinfo {author} {\bibfnamefont {J.-P.}\ \bibnamefont
  {Onnela}}, \bibinfo {author} {\bibfnamefont {A.}~\bibnamefont {Chakraborti}},
  \bibinfo {author} {\bibfnamefont {K.}~\bibnamefont {Kaski}}, \bibinfo
  {author} {\bibfnamefont {J.}~\bibnamefont {Kert\'esz}}, \ and\ \bibinfo
  {author} {\bibfnamefont {A.}~\bibnamefont {Kanto}},\ }\href {\doibase
  10.1103/PhysRevE.68.056110} {\bibfield  {journal} {\bibinfo  {journal} {Phys.
  Rev. E}\ }\textbf {\bibinfo {volume} {68}},\ \bibinfo {pages} {056110}
  (\bibinfo {year} {2003})}\BibitemShut {NoStop}%
\bibitem [{\citenamefont {Jackson}\ and\ \citenamefont
  {Read}(2010{\natexlab{a}})}]{jackson2010a}%
  \BibitemOpen
  \bibfield  {author} {\bibinfo {author} {\bibfnamefont {T.~S.}\ \bibnamefont
  {Jackson}}\ and\ \bibinfo {author} {\bibfnamefont {N.}~\bibnamefont {Read}},\
  }\href {\doibase 10.1103/PhysRevE.81.021130} {\bibfield  {journal} {\bibinfo
  {journal} {Phys. Rev. E}\ }\textbf {\bibinfo {volume} {81}},\ \bibinfo
  {pages} {021130} (\bibinfo {year} {2010}{\natexlab{a}})}\BibitemShut
  {NoStop}%
\bibitem [{\citenamefont {Jackson}\ and\ \citenamefont
  {Read}(2010{\natexlab{b}})}]{jackson2010b}%
  \BibitemOpen
  \bibfield  {author} {\bibinfo {author} {\bibfnamefont {T.~S.}\ \bibnamefont
  {Jackson}}\ and\ \bibinfo {author} {\bibfnamefont {N.}~\bibnamefont {Read}},\
  }\href {\doibase 10.1103/PhysRevE.81.021131} {\bibfield  {journal} {\bibinfo
  {journal} {Phys. Rev. E}\ }\textbf {\bibinfo {volume} {81}},\ \bibinfo
  {pages} {021131} (\bibinfo {year} {2010}{\natexlab{b}})}\BibitemShut
  {NoStop}%
\bibitem [{\citenamefont {Sweeney}\ and\ \citenamefont
  {Middleton}(2013)}]{sweeny2013}%
  \BibitemOpen
  \bibfield  {author} {\bibinfo {author} {\bibfnamefont {S.~M.}\ \bibnamefont
  {Sweeney}}\ and\ \bibinfo {author} {\bibfnamefont {A.~A.}\ \bibnamefont
  {Middleton}},\ }\href {\doibase 10.1103/PhysRevE.88.032129} {\bibfield
  {journal} {\bibinfo  {journal} {Phys. Rev. E}\ }\textbf {\bibinfo {volume}
  {88}},\ \bibinfo {pages} {032129} (\bibinfo {year} {2013})}\BibitemShut
  {NoStop}%
\bibitem [{\citenamefont {Fenner}\ and\ \citenamefont
  {frieze}(1982)}]{fenner1982}%
  \BibitemOpen
  \bibfield  {author} {\bibinfo {author} {\bibfnamefont {T.~I.}\ \bibnamefont
  {Fenner}}\ and\ \bibinfo {author} {\bibfnamefont {A.~M.}\ \bibnamefont
  {frieze}},\ }\href@noop {} {\bibfield  {journal} {\bibinfo  {journal}
  {Combinatorica}\ }\textbf {\bibinfo {volume} {2}},\ \bibinfo {pages} {347}
  (\bibinfo {year} {1982})}\BibitemShut {NoStop}%
\bibitem [{\citenamefont {Frieze}(1985)}]{frieze1985}%
  \BibitemOpen
  \bibfield  {author} {\bibinfo {author} {\bibfnamefont {A.}~\bibnamefont
  {Frieze}},\ }\href {\doibase 10.1016/0166-218X(85)90058-7} {\bibfield
  {journal} {\bibinfo  {journal} {Discrete Applied Mathematics}\ }\textbf
  {\bibinfo {volume} {10}},\ \bibinfo {pages} {47} (\bibinfo {year}
  {1985})}\BibitemShut {NoStop}%
\bibitem [{\citenamefont {McDiarmid}(2000)}]{mcDiarmid1989}%
  \BibitemOpen
  \bibfield  {author} {\bibinfo {author} {\bibfnamefont {C.}~\bibnamefont
  {McDiarmid}},\ }\href@noop {} {\bibfield  {journal} {\bibinfo  {journal}
  {London Math. Soc. Lect. Note Ser.}\ }\textbf {\bibinfo {volume} {141}},\
  \bibinfo {pages} {148} (\bibinfo {year} {2000})}\BibitemShut {NoStop}%
\bibitem [{\citenamefont {Janson}(1995)}]{janson1995}%
  \BibitemOpen
  \bibfield  {author} {\bibinfo {author} {\bibfnamefont {S.}~\bibnamefont
  {Janson}},\ }\href@noop {} {\bibfield  {journal} {\bibinfo  {journal} {Rand.
  Struc. Algor.}\ }\textbf {\bibinfo {volume} {7}},\ \bibinfo {pages} {337}
  (\bibinfo {year} {1995})}\BibitemShut {NoStop}%
\bibitem [{\citenamefont {Flaxman}(2007)}]{flaxman2007}%
  \BibitemOpen
  \bibfield  {author} {\bibinfo {author} {\bibfnamefont {A.~D.}\ \bibnamefont
  {Flaxman}},\ }\href {\doibase 10.37236/1004} {\bibfield  {journal} {\bibinfo
  {journal} {Electr. J. Comb.}\ }\textbf {\bibinfo {volume} {14}},\ \bibinfo
  {pages} {N3} (\bibinfo {year} {2007})}\BibitemShut {NoStop}%
\bibitem [{\citenamefont {Cooper}\ \emph {et~al.}(2016)\citenamefont {Cooper},
  \citenamefont {Frieze}, \citenamefont {Ince}, \citenamefont {Janson},\ and\
  \citenamefont {Spencer}}]{cooper2016}%
  \BibitemOpen
  \bibfield  {author} {\bibinfo {author} {\bibfnamefont {C.}~\bibnamefont
  {Cooper}}, \bibinfo {author} {\bibfnamefont {A.}~\bibnamefont {Frieze}},
  \bibinfo {author} {\bibfnamefont {N.}~\bibnamefont {Ince}}, \bibinfo {author}
  {\bibfnamefont {S.}~\bibnamefont {Janson}}, \ and\ \bibinfo {author}
  {\bibfnamefont {J.}~\bibnamefont {Spencer}},\ }\href {\doibase
  10.1017/S0963548315000024} {\bibfield  {journal} {\bibinfo  {journal} {Comb.
  Prob. Comp.}\ }\textbf {\bibinfo {volume} {25}},\ \bibinfo {pages} {89–107}
  (\bibinfo {year} {2016})}\BibitemShut {NoStop}%
\bibitem [{\citenamefont {Bucklew}(2004)}]{bucklew2004}%
  \BibitemOpen
  \bibfield  {author} {\bibinfo {author} {\bibfnamefont {J.~A.}\ \bibnamefont
  {Bucklew}},\ }\href@noop {} {\emph {\bibinfo {title} {Introduction to rare
  event simulation}}}\ (\bibinfo  {publisher} {Springer-Verlag},\ \bibinfo
  {address} {New York},\ \bibinfo {year} {2004})\BibitemShut {NoStop}%
\bibitem [{\citenamefont {Hartmann}(2011)}]{largest-2011}%
  \BibitemOpen
  \bibfield  {author} {\bibinfo {author} {\bibfnamefont {A.~K.}\ \bibnamefont
  {Hartmann}},\ }\href {\doibase 10.1140/epjb/e2011-10836-4} {\bibfield
  {journal} {\bibinfo  {journal} {Eur. Phys. J. B}\ }\textbf {\bibinfo {volume}
  {84}},\ \bibinfo {pages} {627} (\bibinfo {year} {2011})}\BibitemShut
  {NoStop}%
\bibitem [{\citenamefont {den Hollander}(2000)}]{denHollander2000}%
  \BibitemOpen
  \bibfield  {author} {\bibinfo {author} {\bibfnamefont {F.}~\bibnamefont {den
  Hollander}},\ }\href@noop {} {\emph {\bibinfo {title} {Large Deviations}}}\
  (\bibinfo  {publisher} {American Mathematical Society},\ \bibinfo {address}
  {Providence},\ \bibinfo {year} {2000})\BibitemShut {NoStop}%
\bibitem [{\citenamefont {Touchette}(2009)}]{touchette2009}%
  \BibitemOpen
  \bibfield  {author} {\bibinfo {author} {\bibfnamefont {H.}~\bibnamefont
  {Touchette}},\ }\href {\doibase 10.1016/j.physrep.2009.05.002} {\bibfield
  {journal} {\bibinfo  {journal} {Physics Reports}\ }\textbf {\bibinfo {volume}
  {478}},\ \bibinfo {pages} {1 } (\bibinfo {year} {2009})}\BibitemShut
  {NoStop}%
\bibitem [{\citenamefont {Erd\H{o}s}\ and\ \citenamefont
  {R\'enyi}(1960)}]{erdoes1960}%
  \BibitemOpen
  \bibfield  {author} {\bibinfo {author} {\bibfnamefont {P.}~\bibnamefont
  {Erd\H{o}s}}\ and\ \bibinfo {author} {\bibfnamefont {A.}~\bibnamefont
  {R\'enyi}},\ }\href@noop {} {\bibfield  {journal} {\bibinfo  {journal} {Publ.
  Math. Inst. Hungar. Acad. Sci.}\ }\textbf {\bibinfo {volume} {5}},\ \bibinfo
  {pages} {17} (\bibinfo {year} {1960})}\BibitemShut {NoStop}%
\bibitem [{\citenamefont {Cormen}\ \emph {et~al.}(2001)\citenamefont {Cormen},
  \citenamefont {Clifford}, \citenamefont {Leiserson},\ and\ \citenamefont
  {Rivest}}]{cormen2001}%
  \BibitemOpen
  \bibfield  {author} {\bibinfo {author} {\bibfnamefont {T.~H.}\ \bibnamefont
  {Cormen}}, \bibinfo {author} {\bibfnamefont {S.}~\bibnamefont {Clifford}},
  \bibinfo {author} {\bibfnamefont {C.~E.}\ \bibnamefont {Leiserson}}, \ and\
  \bibinfo {author} {\bibfnamefont {R.~L.}\ \bibnamefont {Rivest}},\
  }\href@noop {} {\emph {\bibinfo {title} {Introduction to Algorithms}}}\
  (\bibinfo  {publisher} {MIT Press},\ \bibinfo {address} {Cambridge (USA)},\
  \bibinfo {year} {2001})\BibitemShut {NoStop}%
\bibitem [{\citenamefont {Hartmann}(2002)}]{align2002}%
  \BibitemOpen
  \bibfield  {author} {\bibinfo {author} {\bibfnamefont {A.~K.}\ \bibnamefont
  {Hartmann}},\ }\href {\doibase 10.1103/PhysRevE.65.056102} {\bibfield
  {journal} {\bibinfo  {journal} {Phys. Rev. E}\ }\textbf {\bibinfo {volume}
  {65}},\ \bibinfo {pages} {056102} (\bibinfo {year} {2002})}\BibitemShut
  {NoStop}%
\bibitem [{\citenamefont {Hartmann}(2025)}]{les_houches2024}%
  \BibitemOpen
  \bibfield  {author} {\bibinfo {author} {\bibfnamefont {A.~K.}\ \bibnamefont
  {Hartmann}},\ }\href {\doibase 10.21468/SciPostPhysLectNotes.100} {\bibfield
  {journal} {\bibinfo  {journal} {SciPost Phys. Lect. Notes}\ ,\ \bibinfo
  {pages} {100}} (\bibinfo {year} {2025})}\BibitemShut {NoStop}%
\bibitem [{\citenamefont {Metropolis}\ \emph {et~al.}(1953)\citenamefont
  {Metropolis}, \citenamefont {Rosenbluth}, \citenamefont {Rosenbluth},
  \citenamefont {Teller},\ and\ \citenamefont {Teller}}]{metropolis1953}%
  \BibitemOpen
  \bibfield  {author} {\bibinfo {author} {\bibfnamefont {N.}~\bibnamefont
  {Metropolis}}, \bibinfo {author} {\bibfnamefont {A.~W.}\ \bibnamefont
  {Rosenbluth}}, \bibinfo {author} {\bibfnamefont {M.~N.}\ \bibnamefont
  {Rosenbluth}}, \bibinfo {author} {\bibfnamefont {A.}~\bibnamefont {Teller}},
  \ and\ \bibinfo {author} {\bibfnamefont {E.}~\bibnamefont {Teller}},\ }\href
  {\doibase 10.1063/1.1699114} {\bibfield  {journal} {\bibinfo  {journal} {J.
  Chem. Phys.}\ }\textbf {\bibinfo {volume} {21}},\ \bibinfo {pages} {1087}
  (\bibinfo {year} {1953})}\BibitemShut {NoStop}%
\bibitem [{\citenamefont {Hastings}(1970)}]{hastings1970}%
  \BibitemOpen
  \bibfield  {author} {\bibinfo {author} {\bibfnamefont {W.~K.}\ \bibnamefont
  {Hastings}},\ }\href {\doibase 10.1093/biomet/57.1.97} {\bibfield  {journal}
  {\bibinfo  {journal} {Biometrika}\ }\textbf {\bibinfo {volume} {57}},\
  \bibinfo {pages} {97} (\bibinfo {year} {1970})}\BibitemShut {NoStop}%
\bibitem [{\citenamefont {Newman}\ and\ \citenamefont
  {Barkema}(1999)}]{newman1999}%
  \BibitemOpen
  \bibfield  {author} {\bibinfo {author} {\bibfnamefont {M.~E.~J.}\
  \bibnamefont {Newman}}\ and\ \bibinfo {author} {\bibfnamefont {G.~T.}\
  \bibnamefont {Barkema}},\ }\href@noop {} {\emph {\bibinfo {title} {Monte
  {C}arlo Methods in Statistical Physics}}}\ (\bibinfo  {publisher} {Clarendon
  Press},\ \bibinfo {address} {Oxford},\ \bibinfo {year} {1999})\BibitemShut
  {NoStop}%
\bibitem [{\citenamefont {Landau}\ and\ \citenamefont
  {Binder}(2000)}]{landau2000}%
  \BibitemOpen
  \bibfield  {author} {\bibinfo {author} {\bibfnamefont {D.~P.}\ \bibnamefont
  {Landau}}\ and\ \bibinfo {author} {\bibfnamefont {K.}~\bibnamefont
  {Binder}},\ }\href@noop {} {\emph {\bibinfo {title} {Monte {C}arlo
  Simulations in Statistical Physics}}}\ (\bibinfo  {publisher} {Cambridge
  University Press},\ \bibinfo {address} {Cambridge},\ \bibinfo {year}
  {2000})\BibitemShut {NoStop}%
\bibitem [{Note1()}]{Note1}%
  \BibitemOpen
  \bibinfo {note} {The data is publicly available in the repository DARE of the
  University of Oldenburg \cite {data_spanning_tree}.}\BibitemShut {Stop}%
\bibitem [{\citenamefont {Steele}(2002)}]{steele2002}%
  \BibitemOpen
  \bibfield  {author} {\bibinfo {author} {\bibfnamefont {J.~M.}\ \bibnamefont
  {Steele}},\ }in\ \href@noop {} {\emph {\bibinfo {booktitle} {Mathematics and
  Computer Science II}}},\ \bibinfo {editor} {edited by\ \bibinfo {editor}
  {\bibfnamefont {B.}~\bibnamefont {Chauvin}}, \bibinfo {editor} {\bibfnamefont
  {P.}~\bibnamefont {Flajolet}}, \bibinfo {editor} {\bibfnamefont
  {D.}~\bibnamefont {Gardy}}, \ and\ \bibinfo {editor} {\bibfnamefont
  {A.}~\bibnamefont {Mokkadem}}}\ (\bibinfo  {publisher} {Birkh{\"a}user
  Basel},\ \bibinfo {address} {Basel},\ \bibinfo {year} {2002})\ pp.\ \bibinfo
  {pages} {223--245}\BibitemShut {NoStop}%
\bibitem [{\citenamefont {Hartmann}\ \emph {et~al.}(2018)\citenamefont
  {Hartmann}, \citenamefont {Doussal}, \citenamefont {Majumdar}, \citenamefont
  {Rosso},\ and\ \citenamefont {Schehr}}]{kpz2018}%
  \BibitemOpen
  \bibfield  {author} {\bibinfo {author} {\bibfnamefont {A.~K.}\ \bibnamefont
  {Hartmann}}, \bibinfo {author} {\bibfnamefont {P.~L.}\ \bibnamefont
  {Doussal}}, \bibinfo {author} {\bibfnamefont {S.~N.}\ \bibnamefont
  {Majumdar}}, \bibinfo {author} {\bibfnamefont {A.}~\bibnamefont {Rosso}}, \
  and\ \bibinfo {author} {\bibfnamefont {G.}~\bibnamefont {Schehr}},\ }\href
  {\doibase 10.1209/0295-5075/121/67004} {\bibfield  {journal} {\bibinfo
  {journal} {Europhys. Lett.}\ }\textbf {\bibinfo {volume} {121}},\ \bibinfo
  {pages} {67004} (\bibinfo {year} {2018})}\BibitemShut {NoStop}%
\bibitem [{\citenamefont {Sarikhani}\ and\ \citenamefont
  {Hartmann}(2025)}]{data_spanning_tree}%
  \BibitemOpen
  \bibfield  {author} {\bibinfo {author} {\bibfnamefont {M.}~\bibnamefont
  {Sarikhani}}\ and\ \bibinfo {author} {\bibfnamefont {A.~K.}\ \bibnamefont
  {Hartmann}},\ }\href {10.57782/BKJRBF} {\enquote {\bibinfo {title} {Data and
  gnuplot plot files},}\ }\bibinfo {howpublished} {DARE public respository,
  DOI: 10.57782/BKJRBF} (\bibinfo {year} {2025})\BibitemShut {NoStop}%
\end{thebibliography}%

\end{document}